# Investigating strain between phase-segregated domains in Cu-deficient $CuInP_2S_6$

Rahul Rao,[1*] Ryan Selhorst,[1,2] Jie Jiang,[1] Benjamin S. Conner,[3,4] Ryan Siebenaller,[5] Emmanuel Rowe,[1,4,6-8] Andrea Giordano,[1,4] Ruth Pachter[1] and Michael A. Susner[1]

[1]Materials and Manufacturing Directorate, Air Force Research Laboratory, Wright-Patterson Air Force Base, OH 45433, USA

[2]UES Inc., Dayton, OH 45433, USA

[3]Sensors Directorate, Air Force Research Laboratory, Wright-Patterson Air Force Base, OH 45433, USA

[4]National Research Council, Washington D.C. 20001, USA

[5]Depatment of Materials Science and Engineering, The Ohio State University, Columbus, OH 43210 USA

[6]Department of Engineering Technology, Middle Tennessee State University, Murfreesboro, TN, 37132 USA

[7]Department of Astronomy and Physics, Vanderbilt University, Nashville, TN, 37235 USA

[8]Department of Life and Physical Sciences, Fisk University, Nashville, TN, 37208 USA

## Abstract

$CuInP_2S_6$ (CIPS) is an emerging layered ferroelectric material with a $T_C$ above room temperature. When synthesized with Cu deficiencies (i.e., $Cu_{1-x}In_{1+x/3}P_2S_6$), the material segregates into CIPS and $In_{4/3}P_2S_6$ (IPS) self-assembled heterostructures within the same single crystal. This segregation results in significant in-plane and out-of-plane strains between the CIPS and IPS phases as the volume fraction of CIPS (IPS) domains shrink (grow) with decreasing Cu fraction. Here, we synthesized CIPS with varying amounts of Cu (x = 0, 0.2, 0.3, 0.4, 0.5, 0.7, 0.8 and 1) and measured the strains between the CIPS and IPS phases through the evolution of the respective Raman, infrared, and optical reflectance spectra. Density functional theory calculations revealed vibrational modes unique to the CIPS and IPS phases, which can be used to distinguish between the two phases through two-dimensional Raman mapping. A comparison of the composition-dependent frequencies and intensities of the CIPS and IPS Raman peaks showed interesting trends with decreasing CIPS phase fraction (i.e., Cu/In ratio). Our data reveal red- and blue-shifted Raman and infrared peak frequencies that we correlate to lattice strains arising from the segregation of the material into CIPS and IPS chemical domains. The strain is highest for a Cu/In ratio of 0.33 ($Cu_{0.4}In_{1.2}P_2S_6$), which we attribute to equal and opposite strains exerted by the CIPS and IPS phases on each other. In addition, bandgaps extracted from the optical reflectance spectra revealed a decrease in values, with the lowest value (~ 2.3 eV) for $Cu_{0.4}In_{1.2}P_2S_6$.

* Correspondence - rahul.rao.2@us.af.mil

## 1. Introduction

$CuInP_2S_6$ (CIPS) is a two-dimensional (2D) room temperature ferroelectric material (Curie temperature $T_C$ ~315 K) that exhibits several interesting properties such as negative piezoelectricity,[1] the electrocaloric effect,[2] giant electrostriction,[3] dielectric tunability,[3] and flexoelectricity.[4,5] Beyond the pure phase material, synthesis with Cu deficiency (i.e. $Cu_{1-x}In_{1+x/3}P_2S_6$) results in the creation of 1) a high temperature cation liquid phase composed of mobile $Cu^+$, $In^{3+}$, and vacant sites above ~520 K, and 2) the formation of self-assembled heterostructures of ferroelectric CIPS and paraelectric $In_{4/3}P_2S_6$ (IPS) phases within the same crystal below ~520 K.[6,7] The interplay between the two different crystal structures leads to an increase in the overall $T_C$ of the ferroelectric CIPS phase (up to ~335 K) as well as interesting physics such as a tunable quadruple potential well for the Cu atomic ordering,[8] large anharmonicity[9] and nonlinear optical effects,[10] and enhanced conductivity at the CIPS/IPS interface.[11]

Previous studies[6,7] have shown that, depending on the amount of Cu present, the phase segregation results in a number of structural and morphological changes. The loss of Cu results in a shrinking of the volume fraction of the CIPS domains and an accompanying increase in the volume fraction of the IPS domains in the material. IPS evinces a smaller layer spacing (defined as the thickness of one lamella and one van der Waals gap) and a smaller area per $P_2S_6$ anion structural sub-unit, giving it a $T_C$-enhancing compressive effect when mated against CIPS domains. This strain effect also serves to stretch the IPS phase when interfaced with the CIPS. The evolution of these relationships is complex (cf. Ref. [7]). As the volume fraction of CIPS domains decreases with decreasing Cu content, the IPS fraction increases and exerts increasing pressure on the CIPS, thus causing the $T_C$ to increase. In addition, the larger phase fraction of the IPS results in a gradual decrease in layer spacing with decreasing Cu.[6] In order to engineer future devices based on Cu-deficient CIPS, it is therefore important to gain further understanding into the structural changes and lattice strains that occur due to the phase segregation.

Here, we study these structural changes through Raman spectroscopy measurements and elucidate the complex relationships between strain and the vibrational modes present in the CIPS and IPS phases, calculated using density functional theory (DFT). Examination of Raman spectra from single-crystalline $Cu_{1-x}In_{1+x/3}P_2S_6$ with varying amounts of Cu reveals unique peaks corresponding to the CIPS and IPS phases; the intensities of these CIPS and IPS peaks exhibit a monotonic decrease and increase, respectively with a reduction in Cu. A detailed analysis of their frequency dependences with respect to decreasing Cu concentration reveals both blueshifts and redshifts, depending on the vibrational mode. These frequency shifts arise due to lattice strain, which, depending on the amount of Cu, can have a significant impact on the structure of the material due to its segregation into CIPS and IPS phases. The lattice strains are also observed to affect the infrared vibrational modes and, along with the Raman mode frequencies, exhibit the maximum deviation from the pure phase mode frequencies for a Cu/In ratio between 0.3 and 0.4. Reflectance spectroscopy measurements show a decrease in the bandgap due to strain and also exhibit a minimum value for Cu/In = 0.33. These results highlight the significant chemical composition-dependent strains within the CIPS lattice and their effect on their optical and electronic properties.

## 2. Experimental

CIPS-IPS single crystals were synthesized using the general procedures outlined in Ref. [12]. Briefly, CIPS with varying amounts of Cu ($Cu_{1-x}In_{1+x/3}P_2S_6$) was synthesized through a self-flux growth technique. The precursor $P_2S_5$ (prepared from Alfa Aesar Puratronic elements, 99.999+% purity, at 300°C for 12 hours in an evacuated fused silica ampoule) was combined with Cu (Alfa Aesar Puratronic, 99.999 % purity, reduced in $H_2$) and In (Alfa Aesar 99.999% purity) in the ratio Cu:In:$P_2S_5$ 1-x:1+x/3:3 in 2 mL $Al_2O_3$ crucibles. The crucibles were arranged in a standard flux growth configuration with the growth crucible on the bottom and an inverted crucible placed directly on top to catch any unreacted flux during centrifugation. We then sealed these crucibles in a fused silica tube under ~250 Torr Ar, heated slowly to 650°C, held at that temperature for 36 hrs, and cooled at a rate of 1.8°C/hr to 350°C when they were centrifuged to decant the flux. Owing to the kinetically-sensitive nature of the heterostructured samples with respect to the formation of their chemical domains,[7,13] after decanting the samples we placed them, inverted, back into the furnace and cooled to room temperature at a rate of 10 $^{o}$C/hr. Sample compositions were determined by subjecting at least three distinct single crystal specimens of each batch to multiple-spot SEM/EDS analysis (9-12 spots total per batch) using a Thermo Scientific Ultra Dry EDS spectrometer joined with a JEOL JSM-6060 SEM. The sample compositions along with their corresponding Cu/In ratios are listed in Table 1.

Following crystal synthesis, we collected micro-Raman spectra in a Renishaw inVia Raman microscope using 514.5 nm excitation. A 100 x objective lens (NA = 0.75) was used to focus onto the samples and the laser power set to under 1 mW to avoid sample heating. 2D Raman maps with a spacing of 600 nm (spot size of our objective lens) were collected across at least three regions from each sample. Individual spectra corresponding to the CIPS and IPS regions within the maps were extracted and averaged to produce the spectra shown in Figs. 1 and 2.

Attenuated total reflectance (ATR) Fourier transform infrared (FTIR) spectra were collected using a Bruker alpha spectrometer equipped with a DTGS detector. Mid-IR spectra were collected between 400 - 3000 cm$^{-1}$ with a spectral resolution of 2 cm$^{-1}$. UV-Vis optical reflectance spectra were collected with a Craic microspectrophotometer. Reflectance spectra between 200 – 1100 nm were collected by focusing the light source (xenon lamp) onto the samples through a 74x magnification objective lens. Over a dozen spectra were collected at different spots from each sample and averaged for analysis. The averaged reflectance spectra were converted into Tauc plots using the Kubelka-Munk function,[14] followed by estimation of the bandgaps by measuring the tangents (in Origin Pro) to the curves (see SI for procedural details). High resolution optical images were collected in the high dynamic range (HDR) mode with a Keyence optical microscope using a 150x magnification objective. The HDR mode captures 16-bit images rather than the typical 8-bit images collected in conventional optical microscopes. To capture these 16-bit images, the camera varies the shutter speed and take multiple images at different brightness levels. Samples that are typically difficult to observe using conventional microscopes, such as low-contrast and reflective surfaces, can be easily imaged in high detail using the HDR mode.

DFT calculations were performed with the Vienna *ab initio* simulation package (VASP 5.4),[15,16] applying the projector augmented-wave (PAW) potential. The Kohn-Sham equations were solved using a plane wave basis set with an energy cutoff of 500 eV. The initial ferroelectric structure of CIPS employed was from the measured ferroelectric bulk structure,[17] optimized using the Perdew-Burke-Ernzerhof (PBE) exchange-correlation functional,[18] including the D3 correction of Grimme,[19] as inclusion of London dispersion was shown to be important.[20,21] A 8×8×3 *k*-point sampling was used in this case. The initial paraelectric IPS bulk structure was taken from the crystal structure of Ref. [6], optimized using PBE+D3 and 8×4×3 *k* sampling. Geometries were fully relaxed regarding lattice parameters and interatomic distances until forces were less than 0.01 eV/Å. The Raman and infrared (IR) spectra of CIPS and IPS were calculated using Phonopy[22] and Phonopy Spectroscopy[23] at the PBE+D3 level. The Phonopy package was employed to calculate the zone-center phonon frequencies and phonon eigenvectors of the DFT optimized structures using the finite-displacement approach. The Raman spectra were simulated by appropriately averaging the Raman activity tensor calculated for each Raman-active phonon eigenmode at the zone center, and the IR spectra by computing the mode dynamical charge associated with each phonon eigen-displacement.

`

## 3. Results and Discussion

CIPS has a monoclinic crystal structure (space group *Cc*) in the ferrielectric state[17] ($T$ < 315 K). Each layer in CIPS consists of an anionic backbone comprising of $S_6$ octahedra, with metal cations and P-P pairs located within (Fig. 1a shows the top and side views of the CIPS lattice). The $S_6$ octahedra and the phosphorus atoms together form a structural backbone comprised of $(P_2S_6)^{4-}$ anion groups that ionically pair with hexagonally arranged and Cu and In cations. In the ferrielectric state, the $Cu^+$ ions predominantly occupy the "up" positions within a trigonal S-S-S position at the octahedral surface representing the boundary with the van der Waals gap; this in turn results in a small displacement of the $In^{3+}$ ions from the center of the octahedra in the opposing direction. This arrangement of the $Cu^+$ and $In^{3+}$ ions is responsible for the electric polarization in CIPS.[24] As mentioned above, when synthesized with Cu deficiencies, the common anion motif enables the material to spontaneously phase separate into CIPS and IPS domains within the same single crystal;[6] the overall composition can be described as the charge-compensated $Cu_{1-x}In_{1+x/3}P_2S_6$. IPS also evinces a monoclinic crystal structure. However, the Cu sites are replaced by $In^{3+}$ ions or ordered vacant sites necessary for charge balance (Fig. 1b); this structure is paraelectric at room-temperature.[25]

Before analyzing the composition-dependent Raman spectra, it is instructive to first focus our attention on the spectra from the pure-phase CIPS and IPS phases. Fig. 1c shows room-temperature Raman spectra from pure-phase CIPS and IPS along with Lorentzian peak fits to the data. The peaks unique to the CIPS and IPS phases are highlighted in red and blue for CIPS and IPS, respectively. To gain insights into the many vibrational modes in CIPS and IPS, we computed Raman spectra for the room temperature bulk structures (Fig. 1d). The calculated vibrational frequencies for CIPS and IPS are underestimated in comparison to the measured Raman spectra due to the level of theory employed,[19,20] which does not consider many-body effects, but trends can be nevertheless be elucidated (see overall agreement of the calculated Raman spectra in Fig. 1d with experimental data in Fig. 1c, as well as Table S1). Our calculated Raman spectra are also in agreement with previous calculations for CIPS.[21] Interestingly, the single peak at 311.9 $cm^{-1}$ in the work by Neal *et al.*[21] is split into two sub-peaks at 316.1 and 294.5 $cm^{-1}$ in our results, consistent with our measurements. Our predicted Raman frequency shifts for peak pairs in pure phase CIPS and IPS (*i.e.* peaks with similar mode vibrations) agree with measurements, for example, the predicted pair at (354.30, 359.70) $cm^{-1}$ with a frequency difference of 5.3 $cm^{-1}$ is consistent with the measured pair of (375.38, 379.29) $cm^{-1}$ with a difference of 3.9 $cm^{-1}$.

While the exact number of peaks between the two pure phase materials differ greatly, in general the Raman spectra exhibit some peaks common to all metal thiophosphate materials.[21,26–39] The general grouping of Raman peaks is as follows: peaks below 100 $cm^{-1}$ correspond to extended vibrational modes involving the metal cations and peaks between 100 and 350 $cm^{-1}$ correspond to deformations of the octahedral cages and involve the S and P atoms. The sharp peak around 380 $cm^{-1}$ is dominated by out-of-plane P-P and S valence vibrations, and the high-frequency modes between 540-620 $cm^{-1}$ correspond mainly to P-P valence vibrations. Phonon eigenvectors of the prominent Raman peaks from CIPS and IPS are shown in Fig. 2 and their computed and experimental peak frequencies, as well as the dominant vibrations are listed in Table S1. Beyond these peaks, we fit the Raman spectra from CIPS and IPS to 21 and 28 peaks, respectively. The peak frequencies and their corresponding vibrational mode descriptions are listed in Table S2. The higher number of Raman peaks in IPS can be attributed to peaks arising from parts of the IPS phase occupied by the $In^{3+}$ ions and peaks from the octahedra with vacant sites.

Fig. 3 shows room-temperature Raman spectra from $Cu_{1-x}In_{1+x/3}P_2S_6$ single crystals with varying amounts of Cu (nominal x = 0, 0.2, 0.3, 0.5, 0.6, 0.7, 0.8 and 1). The nominal sample compositions, and those determined from energy-dispersive x-ray spectroscopy (EDS) analyses along with their Cu/In ratios are listed in. Table 1. For the purpose of visual clarity, the spectra in Fig. 3 have been vertically offset and normalized with respect to the intensity of the peak around 375 $cm^{-1}$. Going from pure phase CIPS at the bottom to pure phase IPS at the top of the waterfall plot in Fig. 3, it is immediately clear that IPS exhibits many more peaks, likely due to the presence of symmetry-breaking ordered vacant sites. With decreasing Cu content, several peaks appear and grow stronger in intensity while some other peaks diminish; these trends generally correspond with increasing (decreasing) phase fractions of IPS (CIPS). Peaks around 100, 320 and 560 $cm^{-1}$ decrease in intensity with a loss of Cu, indicating that these peaks correspond to vibrational modes involving Cu ions. Likewise, peaks around 125, 137, 300 and 580 $cm^{-1}$ increase in intensity with decreasing Cu content, indicating that these peaks correspond to modes involving In cations within the IPS phase. The peaks that decrease and increase in intensity with decreasing Cu and that can be attributed to CIPS and IPS are highlighted with red and blue colors, respectively in Fig. 3. Although the appearance of the Raman peaks correlates with phase fraction, the behaviors of specific vibrational modes are not monotonic in nature, a phenomenon we will describe in more detail below. We note that each spectrum shown in Fig. 3 is an average of several spectra collected through 2D Raman mapping, which is discussed next.

The observation of Raman peaks specific to the CIPS and IPS phases suggests that it should be possible to visualize these domains through a 2D Raman spectral image. The top row in Fig. 4 shows optical microscope images collected from three compositions of $Cu_{1-x}In_{1+x/3}P_2S_6$ (x = 0.8, 0.4 and 0.2). These images were collected in the high dynamic range (HDR) mode in a high-resolution microscope, which offers the necessary contrast to clearly distinguish the CIPS and IPS phases. The striations visible in the optical images in Fig. 4 correspond to the two separate phases, with the light (dark) regions corresponding to CIPS (IPS). The evolution of the structure with decreasing Cu content is noticeably visible in the images, with the CIPS (light regions) decreasing in area fraction owing to programmed Cu deficiency in the pre-reacted mixture and the subsequent increase in the extent of the IPS fraction. As mentioned above, for $Cu_{0.4}In_{1.2}P_2S_6$ the CIPS and IPS domains are nearly equal in size, whereas one of the phases is dominant for the other two compositions. The domain structures in optical images shown in Fig. 4 are similar to those shown in our previous publications,[6,7] where the phases could be distinguished in atomic force microscopy phase and topography images as well as ferroelectric (CIPS) and paraelectric (IPS) domains in piezoresponse force microscopy.

These domains can also be visualized through 2D Raman mapping. The bottom row in Fig. 4 shows corresponding 2D Raman images, where we have plotted the intensity ratios between the CIPS and IPS peaks. The Raman maps in Fig. 4 show the intensity ratios of the low-frequency deformation modes ~103 (139) $cm^{-1}$ in CIPS (IPS); the intensity ratios between the other Raman peaks (Fig. 3) also result in similar images. The Raman maps show that it is quite easy to visualize the phase contrast in CIPS-IPS, with the domain morphologies in the Raman maps appearing identical to those obtained in the optical microscope images (Fig. 4, top row). As the scale bars in Fig. 4 show, the sizes of the CIPS and IPS domains are on the order of microns, therefore our Raman microscope with its spot size of 0.6 nm has enough resolution to distinguish between the two phases. Raman mapping in this way is therefore a quick and easy way to check for segregation of the material into two domains without the need for more time-intensive scanning probe microscopy techniques.

Next, we study the effect of composition and phase morphology on the interplay between the two phases in the self-assembled CIPS-IPS heterostructure, specifically in terms of strain. As mentioned above, previous studies[6] on the effect of Cu content on the structure of CIPS have revealed the following observations:

1. Below ~520 K, the material segregates into CIPS and IPS phases, with the former (latter) decreasing (increasing) in size and lateral extent with decreasing Cu,
2. The smaller layer thickness and area per $P_2S_6$ unit seen in the IPS phase induces a compressive strain on the neighboring CIPS phases, which in turn increases $T_C$, and
3. The interfacial strain effects also impose positive strain on the IPS phase. With these in mind, we next try to unravel the effect of material composition on the Raman peak frequencies and intensities of the CIPS and IPS phases.

We focus on three vibrational modes that are unique to the CIPS and IPS phases, respectively (Fig. 3), and study their frequencies and intensities as a function of decreasing Cu fraction in CIPS. The broad out-of-plane deformation modes around 320 (300) $cm^{-1}$ in CIPS (IPS) can be deconvoluted into at least two peaks, and owing to higher uncertainties in fitting these peaks, we do not include these in the following discussion. Our analysis of the peak frequencies and intensities of the other three peaks is presented in Fig. 5, with frequencies and intensities plotted against the Cu/In ratio, where 1 and 0 correspond to pure-phase CIPS and IPS, respectively. Figs. 5a and 5b show the frequencies (filled data points, left axes) and intensities (hollow data points, right axes) of the low-frequency deformation modes. This low energy mode ~103 (125) $cm^{-1}$ in CIPS (IPS) occurs due out-of-plane rocking motion of the S atoms within the octahedra and out of plane vibrations of the P-P dimers, accompanied by commensurate out-of-plane displacements of the Cu and In ions (Fig. 3). This mode in CIPS decreases in intensity with decreasing Cu/In ratio, with a sharp decline around Cu/In = 0.33 (corresponding to $Cu_{0.4}In_{1.2}P_2S_6$, or x = 0.6 in $Cu_{1-x}In_{1+x/3}P_2S_6$; all compositions are listed in Table 1). At the same time, the peak exhibits a blueshifted frequency with decreasing Cu content. The corresponding mode in IPS (Fig. 5b) increases in intensity and redshifts with decreasing Cu. The frequencies of both peaks vary sharply below Cu/In = 0.33, an effect that we also observe in other peaks and is discussed further below. The contrasting frequency behaviors of the low-energy deformation mode in CIPS and IPS highlight the opposing structural changes occurring within the material as it segregates into separate CIPS and IPS phases due to the removal of Cu ions.

Raman peak blueshifts and redshifts are typically attributed to compressive and tensile strain, respectively [40]. Compressive strains cause bond shortening, and with the force constant remaining the same, this results in a shift of the vibrational modes to higher frequencies. Conversely, bond lengthening by tensile strain results in a lowering of vibrational mode frequencies. The blueshift of the deformation mode in CIPS (especially above Cu/In = 0.33, Fig. 5a) can thus be attributed to compressive strain on the CIPS phase due to the increasing quantity of IPS phase present as $x$ is increased in $Cu_{1-x}In_{1+x/3}P_2S_6$. Since this mode involves out-of-plane displacements of the S and P atoms, a decrease in layer spacing due to the segregation of the material into the two separate CIPS and IPS phases causes intra-layer compression and hence, blueshifted Raman peak frequencies. The larger layer thicknesses and area per $P_2S_6$ subunit observed in the CIPS phase in turn impart a tensile strain on the IPS phase when mated together, causing its deformation mode to redshift (Fig. 5b). As the volume fraction of the IPS phase approaches unity, this redshift diminishes and then disappears entirely for pure-phase IPS.

The high intensity peak ~375-379 $cm^{-1}$, while appearing to be a single mode peak, can actually be deconvoluted into two peaks where the lower and higher frequency peaks are assigned to the CIPS (375 $cm^{-1}$) and IPS (379 $cm^{-1}$), respectively. This is evident in the offset between the peaks from pure-phase CIPS and IPS as revealed by our Lorentzian peak fitting (Fig. 2). This mode occurs mainly due to out-of-plane valence vibrations of the P-P and in-plane lattice vibrations of S atoms (P-S stretch, Fig. 3). Similar to the two peaks described above, the intensity of this peak decreases in the CIPS Raman spectrum with decreasing amount of Cu, while it increases in the IPS spectrum (Figs. 5c and 5d). The frequencies of the P-S stretch mode in both phases exhibit similar trends - they blueshift with decreasing Cu until a Cu/In ratio of ~0.4, after which they do not exhibit any further significant shifts, at least within experimental error. The blueshift of the peak frequencies can be attributed to the decrease in the layer spacing (and consequently P-P bond shortening) associated with in-plane heterostructure formation. At lower amounts of Cu (*i.e.*, Cu/In < 0.4), the compressive strain is likely offset by the tensile strain due to P-S bond lengthening, resulting in the peak frequencies remaining mostly constant.

The last peak we discuss is the high-frequency mode above 550 $cm^{-1}$. This peak appears at ~558 (580) $cm^{-1}$ in CIPS (IPS). Similar to the P-S stretching mode, these modes involve out-of-plane P-P valence vibrations and in-plane S valence vibrations, with greater displacements of the P atoms compared to the S atoms (P-P stretch). While their composition-dependent intensities exhibit the same trends as the other Raman peaks (decreasing and increasing with decreasing Cu content for CIPS and IPS phases, respectively, Figs. 5e and 5f), their frequencies exhibit contrasting behaviors compared to those of the P-S stretching modes. The P-P stretch mode in the CIPS phase remains relatively constant with decreasing Cu content down to a Cu/In ratio of 0.25, below which it redshifts. In the IPS phase, this peak exhibits a monotonic redshift with decreasing Cu content. The frequency trend in CIPS (Fig. 5e) can be attributed to both P-S bond lengthening due to the change in lattice dynamics attributed to the heterostructure formation and P-P bond shortening (due to a decrease in layer spacing), which combined together result in a near-constant peak frequency. As the amount of Cu decreases below a Cu/In ratio of 0.25, the additional strain imposed on the CIPS phase by the growing IPS domains might be the reason why this peak exhibits a redshift for lower Cu concentrations (Fig. 5f). The monotonic redshift of the P-P stretching mode in IPS can be attributed to the greater disparity in P-S bond lengths in IPS compared to CIPS (Fig. S1), with half of the P-S bonds longer than the P-S bonds in CIPS and the other half shorter. This disparity in P-S bond lengths is likely responsible for the observed monotonic redshift of this peak with decreasing amount of Cu. We note that this trend is similar to the behavior of the lowest energy deformation mode ~125 $cm^{-1}$ (up to Cu/In = 0.33 as discussed above and shown in Fig. 5b), which also involves out-of-plane displacements of the P and S atoms, similar to the P-S stretching mode. However, the low-energy deformation mode also involves vibrations of the In ions, which are much larger than the S and P atoms.

As previously mentioned, many of the Raman peaks in the CIPS and IPS phases exhibit an apparent inflection (*i.e.,* sharp redshifts or blueshifts) between a Cu/In ratio of 0.3 and 0.5. The reason for this inflection is unclear at this time, although the images of the composition-dependent phase segregations may provide clues. As can be seen in the first image in Fig. 4, a decrease in Cu fraction to 0.8 results in the formation of long and thin IPS domains. These domains grow in size with decreasing Cu content, and are roughly equal in size for Cu/In = 0.33 ($Cu_{0.4}In_{1.2}P_2S_6$). For the lowest amount of Cu ($Cu_{0.2}In_{1.267}P_2S_6$), the CIPS phase appears as thin ribbons within the majority IPS phase. Considering that the CIPS phase is compressed both in-plane and along the stacking direction,[6,7] it is plausible that, at the appropriate Cu/In ratio, the shrinking (growing) CIPS (IPS) domains exert equal amounts of in-plane and out-of-plane strains on each other, causing the observed inflection in the Raman frequencies shown in Fig. 5. This is also reflected in a significant increase in the number of Raman peaks for compositions below Cu/In = 0.33 (Fig. 1).

We also measured composition-dependent FTIR spectra, which show strains consistent with the Raman data. Fig. 6a shows FTIR transmittance curves for $Cu_{1-x}In_{1+x/3}P_2S_6$ crystals with varying amounts of Cu. Similar to the Raman spectra, the transmittance spectra for Cu/In ratios below 0.4 exhibit more peaks. Note that, for the FTIR spectra, we were unable to collect data through 2D mapping with a microscope, thus the data from each composition shown in Fig. 6a correspond to peaks from both CIPS and IPS phases. We fit these spectra to Lorentzian peaks and show the frequencies of two high-frequency modes as a function of Cu/In ratio in Fig. 6b. These peaks are attributed to in-plane P-S vibrations and both peaks redshift with decreasing Cu fraction down to Cu/In = 0.4, below which they blueshift. This inflection around Cu/In = 0.4 is similar to those observed in the Raman peak frequencies.  Note that comparing our calculated IR spectra for CIPS and IPS (our results for CIPS are in excellent agreement with previous work,[21] see Table S3) show an increase in the number of peaks for IPS (Fig. S2(a)), as measured experimentally (Fig. 5a), for example the split of the peak at ca. 550 $cm^{-1}$. The phonon eigenvectors for peaks with frequencies larger than 500 $cm^{-1}$ show dominant P-S vibrations for both CIPS and IPS (Fig. S2(b)).

Finally, we measured UV-vis reflectance spectra from the samples, which show a dip between 400 – 500 nm due to absorption across the direct bandgap (Figs. S3 and S4). By converting the reflectance curves to Tauc plots using the Kubelka-Munk function[14] (Figs. S3 and S4), and by estimating the tangents to the curves, we extracted the bandgaps. We plot these against the Cu/In ratio in Fig. 6c. Pure phase CIPS exhibits a bandgap around 2.7 eV, which agrees with previously measured values.[41,42] The bandgap reduces with decreasing Cu fraction, with the lowest value approaching ~2.3 eV for Cu/In = 0.33, beyond which it increases again. This inflection in the bandgaps is similar to what we observe in the Raman and IR peak frequencies. A decrease in the bandgap has been observed previously in similar materials upon the application of hydrostatic pressure.[43] It was also previously shown that hydrostatic pressure results in an increase in the Curie temperature in CIPS (~0.1 GPa results in a $T_c$ of ~340 K),[44] and we know that $T_c$ increases in Cu-deficient CIPS.[6] Our previous high pressure Raman spectroscopy study also showed that hydrostatic compression of ~0.1 GPa results in Raman peak blueshifts (e.g. ~ 0.23 $cm^{-1}$ for the high-intensity P-S stretching mode ~378 $cm^{-1}$. Such a blueshift matches well with our observed shifts in Raman peak frequencies with changing Cu fraction in CIPS.

Taken together, our Raman, FTIR and optical spectroscopy studies reveal a complex strain state caused by the phase segregation in CIPS due to decreasing amount of Cu. The differences in the sizes between the growing and shrinking CIPS and IPS domains result in compressive and tensile stress, as well as a decrease in the layer spacing. The pressure exerted by the two phases results in an increase in the Curie temperature, and the complex strain states are also likely responsible for the more exotic properties observed in Cu-deficient CIPS, such as a quadruple well ferroelectric potential. Other characterization methods such as synchrotron diffraction or transmission electron microscopy studies might provide more insights into the microscopic details of the composition-dependent strains between the CIPS and IPS phases, although owing to the structural complexity of CIPS-IPS, Rietveld refinements of a two-phase composition-dependent dataset is not straightforward, and these measurements are beyond the scope of this study.

## 4. Summary

Cu-deficient CIPS self-segregates into separate CIPS and IPS domains, whose sizes and lateral extents depend on the material composition. Here we study several compositions of $Cu_{1-x}In_{1+x/3}P_2S_6$ and reveal Raman peaks unique to the CIPS and IPS phases. A systematic study of their Raman and infrared peak frequencies and intensities with varying Cu content reveals several strain-dependencies, which are the direct result of changes in bond lengths and lattice expansion/contraction due to the self-segregation process. These strains also result in a decrease in the bandgap of the material, with an inflection point observed in the data for a Cu/In ratio between 0.3 and 0.5. The microstructure of the material for this composition consists of nearly equal sized CIPS and IPS domains, which apparently exert equal and opposite strains on each other. Our study thus provides insights into the complex interplay between the heterostructured CIPS and IPS phases and the evolution of these phases as composition is varied. These insights will help in strain-engineering of the ferroelectric properties[4,5] by tuning the CIPS composition and expand the breadth of device applications based on this unique and highly interesting layered ferroelectric material.

**Acknowledgements**

This research was funded by the Air Force Office of Scientific Research (AFOSR) grant LRIR 23RXCOR003 and AOARD-MOST Grant Number F4GGA21207H002. We also acknowledge support from the National Research Council's Senior NRC Associateship program sponsored by the National Academies of Sciences, Engineering, and Medicine.

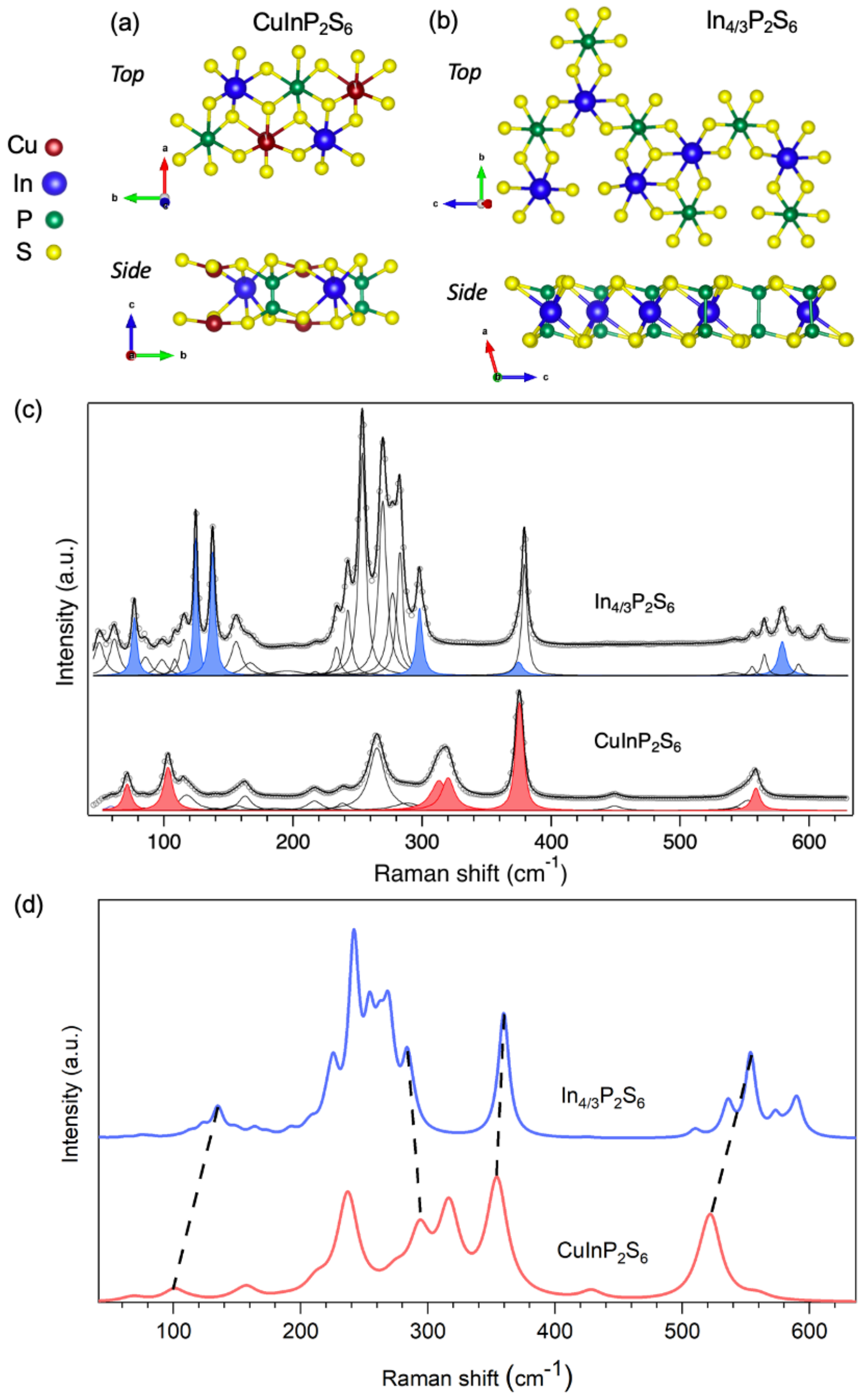

Fig. 1. Schematics of the atomic structures of a monolayer of (a) $CuInP_2S_6$ (CIPS) and (b) $In_{4/3}P_2S_6$ (IPS). The top and side views of the lattices are shown. (c) Raman spectra from CIPS and IPS with Lorentzian peak fits to the spectra. The overall fits are overlaid on top of the raw data. The peaks specific to the CIPS and IPS phases, as discussed in the text, are highlighted by the red and blue peaks, respectively. (d) Calculated Raman spectra for CIPS and IPS. Pairs of peaks in CIPS and IPS are connected by dashed lines.

Table 1. List of sample compositions and Cu/In ratios.

| **Nominal composition ($Cu_{1-x}In_{1+x/3}P_2S_6$)** | ***x*** | **Cu/In ratio (nominal)** | **EDS Composition** | **Cu/In ratio (EDS)** |
|---|---|---|---|---|
| **$CuInP_2S_6$** | 0 | 1 | - | 1 |
| **$Cu_{0.8}In_{1.067}P_2S_6$** | 0.2 | 0.75 | $Cu_{0.76(3)}In_{1.20(3)}P_2S_6$ | 0.63 |
| **$Cu_{0.7}In_{1.1}P_2S_6$** | 0.3 | 0.64 | $Cu_{0.66(3)}In_{1.15(4)}P_2S_6$ | 0.57 |
| **$Cu_{0.5}In_{1.16}P_2S_6$** | 0.5 | 0.43 | $Cu_{0.59(9)}In_{1.18(4)}P_2S_6$ | 0.5 |
| **$Cu_{0.4}In_{1.2}P_2S_6$** | 0.6 | 0.33 | $Cu_{0.29(3)}In_{1.42(3)}P_2S_6$ | 0.2 |
| **$Cu_{0.3}In_{1.233}P_2S_6$** | 0.7 | 0.243 | $Cu_{0.35(4)}In_{1.39(4)}P_2S_6$ | 0.25 |
| **$Cu_{0.2}In_{1.267}P_2S_6$** | 0.8 | 0.158 | $Cu_{0.21(2)}In_{1.44(4)}P_2S_6$ | 0.15 |
| **$In_{4/3}P_2S_6$** | 1 | 0 | - | 0 |

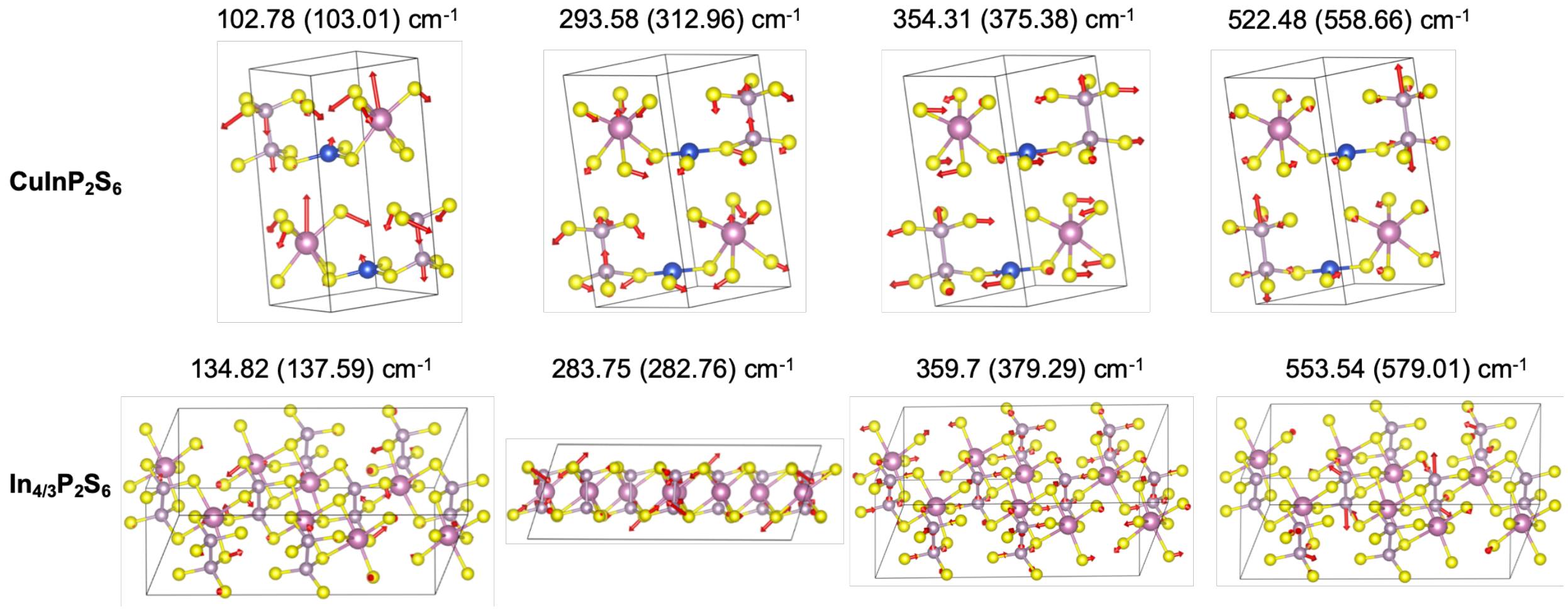

Fig. 2. Phonon eigenvectors for vibrational modes common to CIPS and IPS as highlighted in Figs. 1 and 2. The top and bottom rows depict the vibrational modes in CIPS and IPS, respectively.

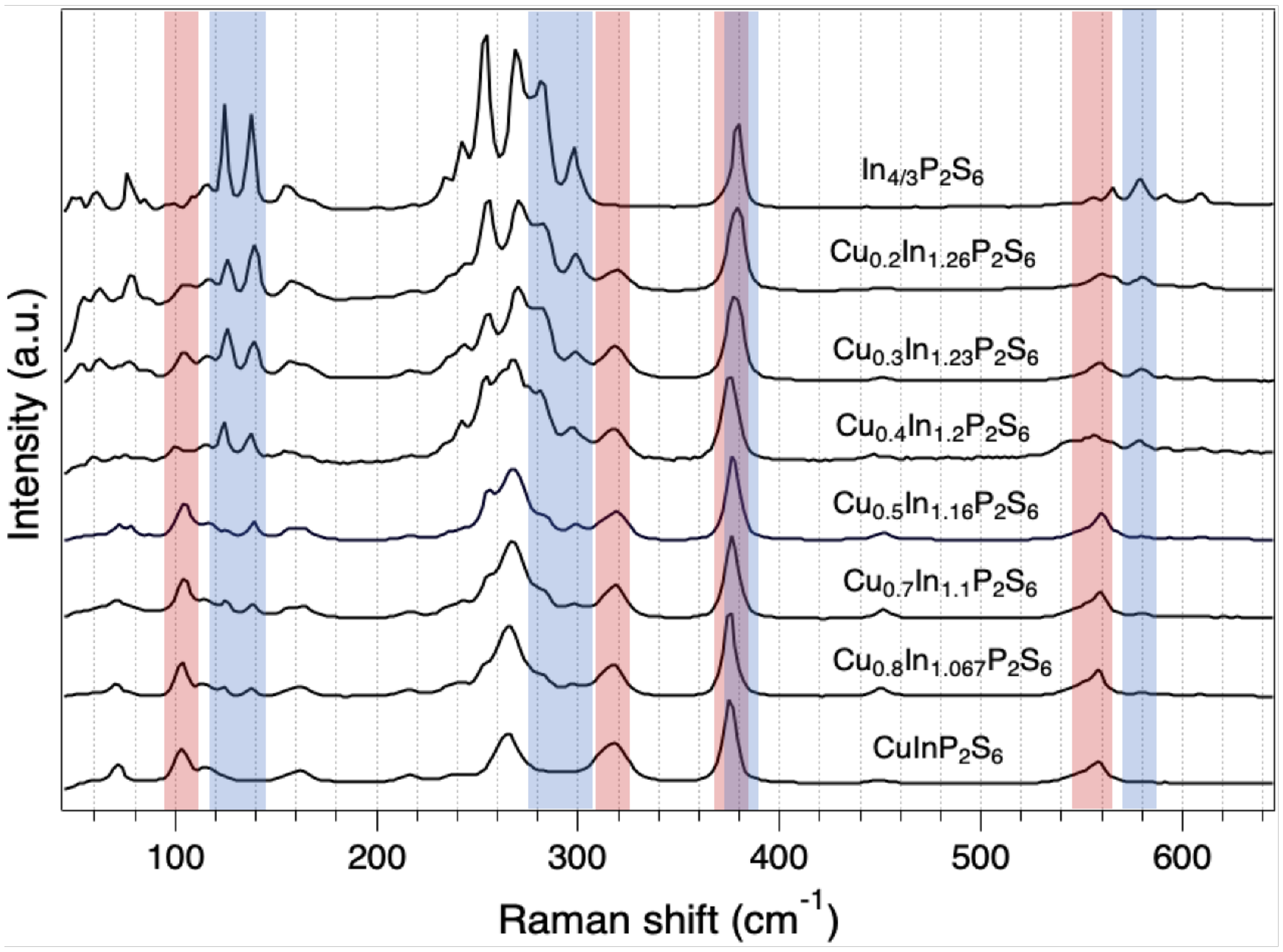

Fig. 3. Room-temperature Raman spectra from $Cu_{1-x}In_{1+x/3}P_2S_6$ with varying amounts of Cu. The nominal peaks specific to the CIPS and IPS phases, as discussed in the text, are highlighted by the red and blue colored bands, respectively.

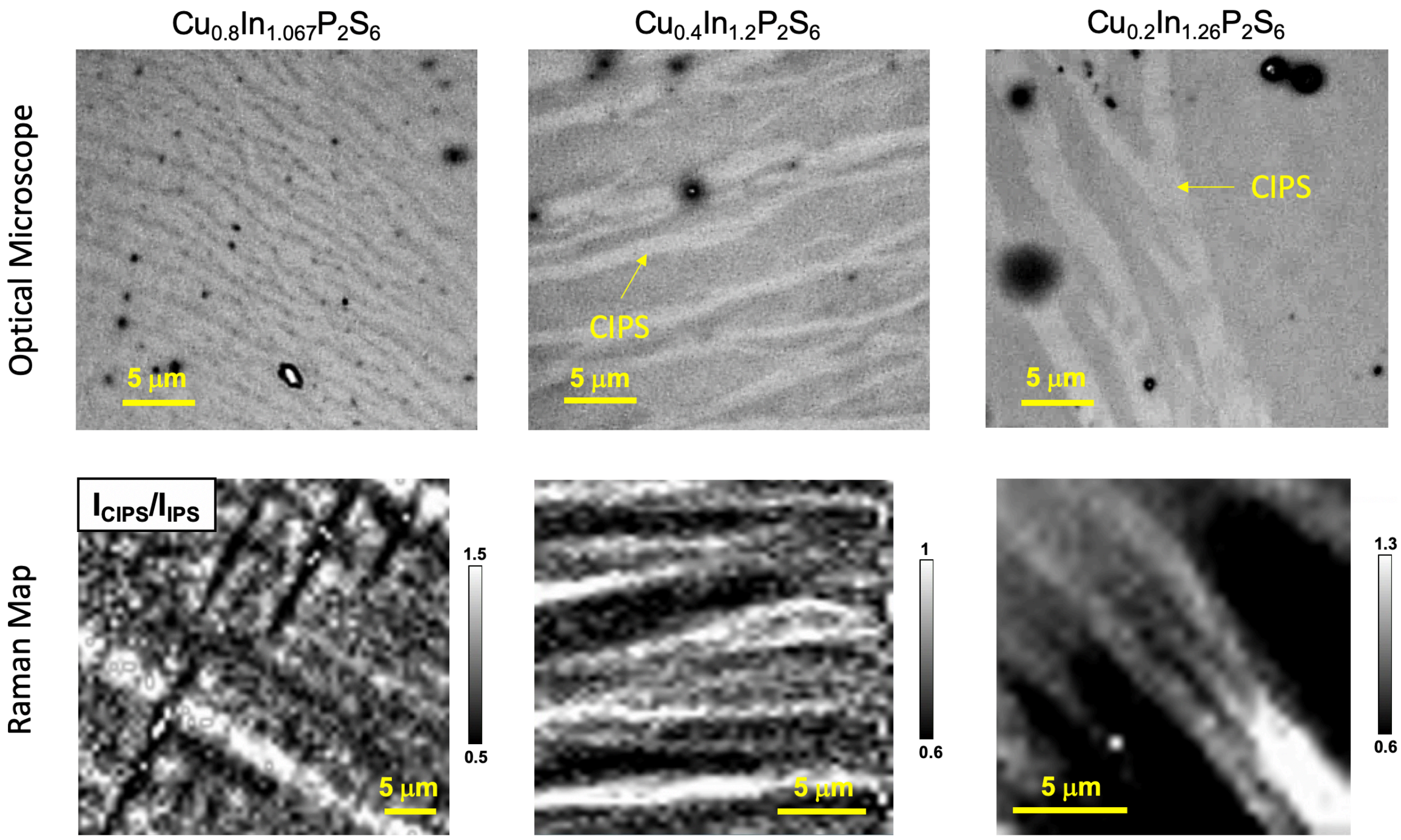

Fig. 4. Top row - Optical microscope images collected from three nominal compositions of $Cu_{1-x}In_{1+x/3}P_2S_6$. The light (dark) regions correspond to the CIPS (IPS) phases. Bottom row - Images generated from 2D Raman maps.

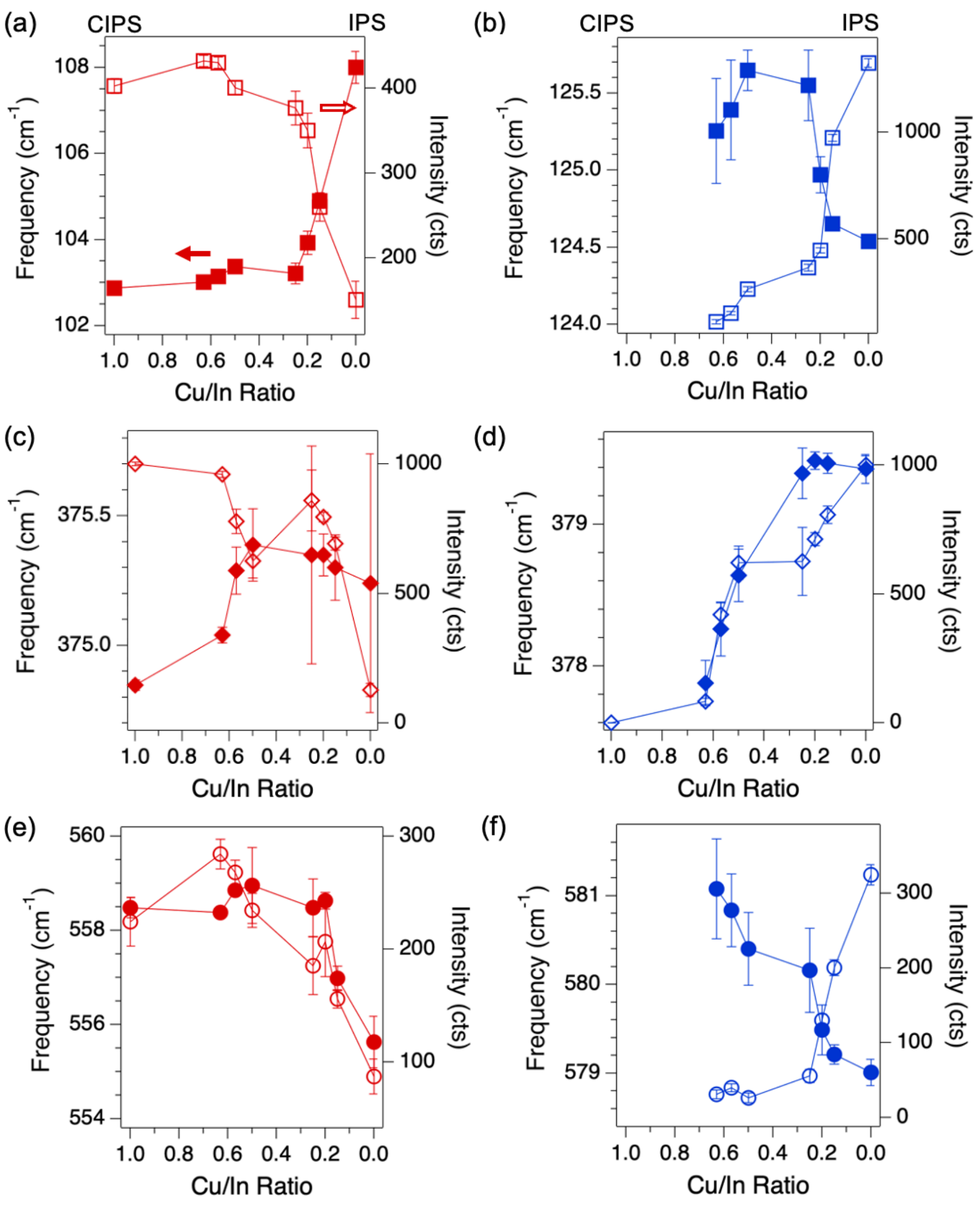

Fig. 5. Composition-dependent Raman peak frequencies and intensities. (a), (c), (e) and (g) Frequencies (filed data points, left axes) and intensities (hollow data points, right axes) as a function of Cu/In ratio (obtained from EDS analysis) for peaks from the CIPS phase. (b), (d), (f) and (h) Frequencies as a function of Cu/In ratio for peaks from the IPS phase. Cu/In ratios of 1 and 0 correspond to pure-phase CIPS and IPS, respectively. The atomic displacements for the vibrational modes are shown schematically in the insets.

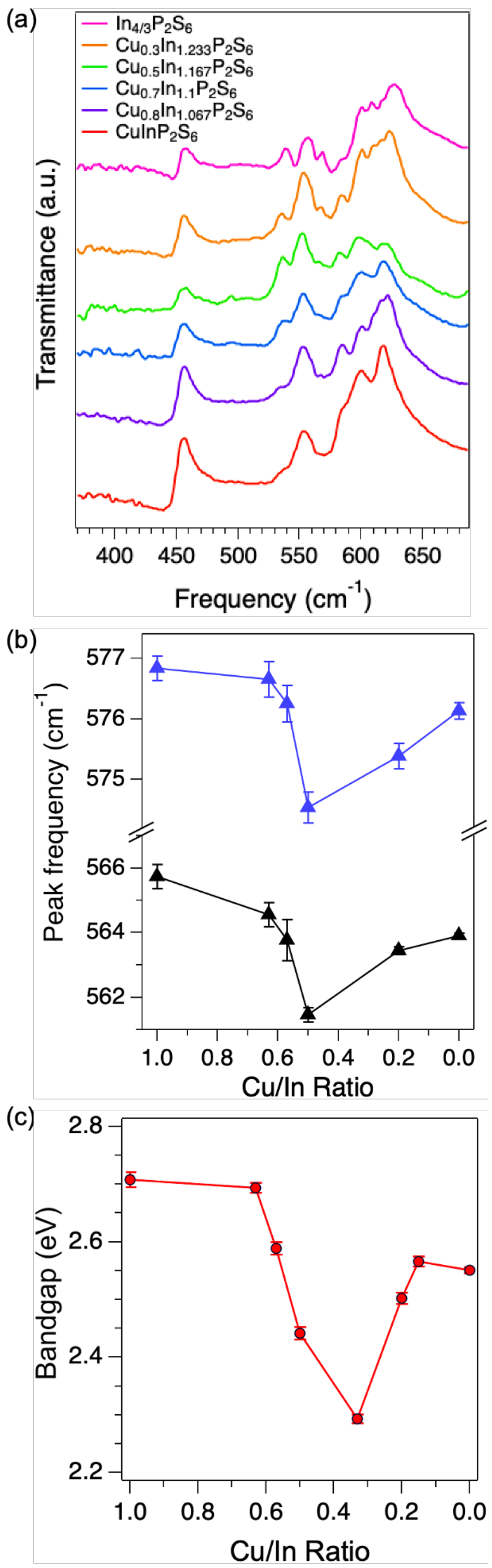

Fig. 6. (a) FTIR transmittance spectra from CIPS-IPS with varying amounts of Cu. (b) Frequencies of two high frequency peaks vs. Cu/In ratio. (c) Bandgap obtained from reflectance spectra vs. Cu/In ratio.